\documentclass[twocolumn,showpacs]{revtex4}

\input epsfig.sty
\usepackage{graphicx}
\usepackage{dcolumn}

\def\be{\begin{equation}}
\def\ee{\end{equation}}

\begin{document}
\title{Cooling of a New Born Compact Star with QCD Phase Transition} 
\author{K.~-W.~Wong}
 \email{kwwong@virginia.edu}
\affiliation{Department of Physics, The Chinese University of Hong
Kong, Shatin, New Territories, Hong Kong, China\\ and Department of 
Astronomy, University of Virginia, PO Box 3818, Charlottesville,VA~
22903, USA}
\author{M.~-C.~Chu}
 \email{mcchu@phy.cuhk.edu.hk}
\affiliation{Department of Physics, The Chinese University of Hong
Kong, Shatin, New Territories, Hong Kong, China}

\date{\today}
 

\begin{abstract}

We study the cooling behaviour of an isolated strange quark star, using an 
equation of state derived from perturbative QCD up to second order in strong
coupling constant, and we compare it with that of a neutron star.  After an
initial rapid cooling, a quark star may undergo the QCD phase
transition to become a neutron star.  
We propose several signatures for such a scenario: a large amount of 
energy can be released due to latent heat, a long duration $\gamma$-ray 
source, and a second neutrino burst after a supernova explosion.

\end{abstract} 


\maketitle



\section{Introduction}
 
The Witten proposal that symmetric deconfined u,d,s-quark matter 
may be the absolute ground state of matter \cite{Witten} has aroused much 
interest, and the properties of strange stars have been widely studied 
since then.  An important question is whether
the observed compact stars are neutron stars or strange stars,
which are made up of deconfined u,d,s quarks.  With the launching of
the new generation of X-ray detectors $Chandra$ and $XMM$,
it is becoming possible now to have accurate measurement of the
radii and surface temperature of compact stars
\cite{Slane,Tsuruta2002,Drake2002,Drake2003}.  However, 
theoretical calculations using the MIT Bag Model equation of state (EOS) 
show that the mass and size of a strange 
star are comparable to those of a neutron star \cite{Alcock,Glendenning}.  
Hence it is important to identify other observables that can be used to
distinguish a strange star from neutron star.  One possibility is to study
the cooling of an isolated compact star. The cooling curves of 
quark matter and neutron matter are found to be significantly different due to 
the difference in their thermal properties and energy 
loss mechanisms \cite{Glendenning,Ng,Usov1998,Usov2001,Usov2002}. 

In this paper, we study the effects of the QCD phase transition on the
cooling of a compact star and possible signatures of the quark phase.
Regardless of the validity of Witten's proposal, 
the formation of quark-gluon plasma should be favoured in high 
temperature and density \cite{shuryak}; 
we therefore suggest that a strange quark star may be formed just 
after a supernova explosion, in which both conditions may be satisfied 
\cite{kawa1}. 
Because the initial temperature is so high \cite{supernova} $T_i \sim 
40$ MeV, the initial compact star is likely to be a bare strange star 
\cite{Usov2001}. 
When it cools 
down to the phase transition temperature $T_p$, the quark matter may
become energetically unstable compared to nuclear matter,  and 
the strange star will convert to a neutron star. During the phase transition, 
a large amount of latent heat, of the order $10^{53}$ erg, can be 
released, which can be a possible energy source of Gamma Ray Bursts (GRB's). 

The latest lattice QCD calculations of $T_p$ \cite{fodor,allton} 
indicate, though 
with relatively large uncertainties at high chemical potential, that
$T_p $ drops from its zero density value of 140 MeV to about
50 MeV at 1.5 times nuclear matter density
$\rho_0=0.17 \textrm{ fm}^{-3}$ and down to a few MeV for
density a few times $\rho_0$.  Some previous proto-neutron star evolution
calculations indeed show that it is feasible to reach the phase transition
in supernovae \cite{Benvenuto}.  While there are still large uncertainties
in both high density QCD and the proto-neutron star evolution, we believe
it is worthwhile to study the possible consequences of the QCD phase
transition in supernovae.
We assume a constant $T_p$ in density throughout the star and present 
results for
$T_p = 1, \ 10$ MeV for comparison.  We adopt the simple picture that
matter at temperature above (below) $T_p$ is in the quark (hadronic)
phase.

It has been argued that the rapid cooling of strange stars by pion 
and $e^-e^+$ pair production can be a power source of 
GRB's \cite{Ng,Usov1998,Usov2001,Usov2002}. 
However, if the QCD phase transition is not considered, the temperature of 
the compact star drops rapidly and the duration of the burst -- being less than
$10^{-2}$ s -- is too short to account for long duration GRB's. 
In our model, the star stays at the phase transition temperature for
a relatively long time, and the photon
luminosity is maintained in the range $10^{48}-10^{54} 
~\textrm{erg s}^{-1}$ with duration $10^{-3}$~s up to $10^4$~s (or even 
longer), which is consistent 
with both long and short GRB's. Thus the latent heat may solve the 
problem of the energy supply of GRB's.

Another signature of quark stars that has been discussed is the relatively 
low luminosity of $\gamma$-rays when the quark star cools 
down \cite{Usov2002}. Our model shows that the luminosity in the 
$\gamma$-ray range can be as high as $10^{38-39}~\textrm{erg s}^{-1}$with 
a very long duration of $\sim 10^{13-14}$ s, which is much longer than the 
case considered by Page and Usov \cite{Usov2002}. The satellite $Integral$ 
launched 
recently is just sensitive to $\gamma$-rays in the energy range of interest. 
If a long duration $\gamma$-ray source is detected, it could be 
a signature of the phase transition \cite{lebrun,oaknin}.

A remarkable feature of a strange star to neutron star phase transition 
is the emission of a second 
neutrino burst after the supernova explosion. A similar 
scenario was also proposed in Aguilera $et$ $al.$'s work \cite{Aguilera},
but with a different physical mechanism. 
In our model, the burst 
is due to the phase transition from a quark star to a neutron star, which
has a higher neutrino emissivity. In Aguilera $et$ $al.$'s work, 
the burst is 
due to the trapping of neutrinos when the temperature is high, which
are then released suddenly when the quark star cools. Nevertheless, both works 
suggest that a second neutrino burst is a signature of the existence 
of quark stars. 

This paper is organised as follows. Section II describes the 
EOS used in the quark phase. The stability of strange quark star is 
investigated in Section III. 
The effect of strange quark mass is studied 
in Section IV. We describe how to calculate the cooling history and phase 
transition of compact stars in Section V and VI. Section VII describes 
the models we used. The calculated results for the various models are
presented in Section VIII.  Section IX is a short discussion and summary 
of our work.

\section{COLD EQUATION OF STATE FROM PERTURBATIVE QCD}
Various EOS's have been used to study the properties of 
strange stars. The most widely used one is the MIT Bag Model due to its 
simple analytic form \cite{Alcock}. It has been pointed out that it is 
difficult to distinguish the strange stars described by the MIT Bag Model 
from neutron stars due to the similiarities in their maximum mass $\sim 2 
M_{\odot}$ ($M_{\odot}$ is a solar mass) and radius 
$\sim 10$ km \cite{Fraga}. On the other hand, Fraga $et$ $al.~$ \cite{Fraga}
studied quark star structure by using the EOS of perturbative QCD for 
cold, dense quark matter up to order $ \alpha_s^2$, using a modern 
determination of the running of the coupling constant. Their results show 
that strange stars can have a radius of about $5.8$ km and a mass of a 
typical neutron star, and such a small compact star can actually be 
distinguished from neutron stars. Since the strong coupling constant
becomes small in the high density limit, perturbative QCD may be a
fair description of matter in the interior of a compact star. Hence we 
follow Fraga's work and examine in details the conditions of 
absolute and global stability of strange quark stars, as well as the 
effects of strange quark mass. It has been argued that perturbative QCD 
may fail to describe the matter at the surface of a strange star due to 
the strong coupling at low density. However, the main structure of a 
compact star is determined by matter properties in the high density regime. 
Moreover, it is worthwhile to study the dependence of compact star 
properties on models of EOS.

The chemical equilibrium of strange quark matter is maintained by the 
weak-interaction reactions: 
\begin{equation}
d\leftrightarrow u+e+\overline{\nu}_e,
\end{equation}
\begin{equation}
s\leftrightarrow u+e+\overline{\nu}_e
\end{equation}
and
\begin{equation}
s+u\leftrightarrow u+d.
\end{equation}
Given the thermodynamic potential of each species $\Omega_i(i=u,d,s,e)$, 
the number densities can be obtained from the thermodynamic relation:
\begin{equation}
n_i=-\frac{\partial \Omega_i}{\partial \mu_i},
\end{equation}
where $\mu_i$ is the chemical potential.
The conditions of chemical equilibrium are:
\begin{equation}
\mu_d=\mu_u+\mu_e,
\end{equation}
\begin{equation}
\mu_d=\mu_s.
\end{equation}
Together with the charge neutrality condition:
\begin{equation}
\frac{2}{3}n_u-\frac{1}{3}(n_d+n_s)-n_e=0,
\end{equation}
the thermodynamic properties will be determined by one independent choice 
of chemical potential only, which we have chosen to be 
$\mu\equiv\mu_d=\mu_s$ in our calculations. The total pressure is 
given by: 
\begin{equation}
P(\mu)=-\sum_i \Omega_i(\mu),
\end{equation}
and the total energy density is:
\begin{equation}
\epsilon(\mu)=\sum_i[\Omega_i(\mu)+\mu_i(\mu)n_i(\mu)].
\end{equation}

At the zero quark mass limit, 
$\Omega_u=\Omega_d=\Omega_s$ implying $n_u=n_d=n_s$. Hence the charge 
neutrality condition is automatically satisfied, without any need of 
electrons.
The perturbative QCD thermodynamic potential at zero temperature has been 
calculated up to order 
$\alpha_s^2$ \cite{Freedman,Baluni}, which in the modified 
minimal subtraction scheme \cite{Baluni} is:
\begin{widetext}
\begin{equation}
\Omega(\mu)=- \frac{N_f\mu^4}{4 \pi^2} \left\{1-2\left( 
\frac{\alpha_s}{\pi}\right)-
\left[G+N_f\ln\left(\frac{\alpha_s}{\pi}\right)+
\left(11-\frac{2}{3}N_f\right)\ln
\frac{\Lambda}{\mu}\right]\left(\frac{\alpha_s}{\pi}\right)^2\right\}, 
\end{equation}
\end{widetext}
where $G={G}_{0}-0.536{N}_{f}+{N}_{f}\ln {N}_{f}$,
${G}_{0}=10.374\pm 0.13$, $N_f$ is the number of quark flavors, 
$\Lambda$ is the renormalization subtraction point, and

\begin{eqnarray}
\alpha_s (\Lambda) & = & \frac{4\pi}{{\beta}_{0}u}
\left[1-\frac{2{\beta}_{1}}{{\beta}_{0}^2}
\frac{\ln (u)}{u} + \right.\nonumber\\
 & & \left. \frac{4{\beta}_{1}^2}{{\beta}_{0}^4u^2}
\left(\left(\ln (u)-\frac{1}{2}\right)^2+
\frac{ {\beta}_{2}{\beta}_{0} }{8 {\beta}_{1}^2}
-\frac{5}{4}\right)\right],
\end{eqnarray}
with $u=\ln (\Lambda^2/\Lambda_{\overline{MS}}^2)$, 
${\beta}_{0}=11-2{N}_{f}/3$,
${\beta}_{1}=51-19{N}_{f}/3$ and
${\beta}_{2}=2857-325{N}_{f}^2/27$. 
The boundary condition of ${\alpha}_{s}=0.3089$ at 
$\Lambda=2$ GeV for ${N}_{f}=3$ gives 
${\Lambda}_{\overline{MS}}=365$ MeV.

\begin{figure}[h]
\epsfig{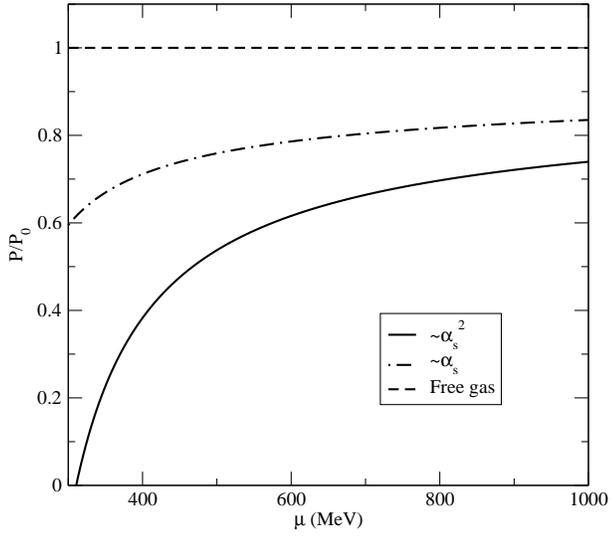}
\caption{The total pressure of strange quark matter, $P$, relative to 
the 
pressure of an ideal gas, $P_0$, as a function of chemical potential, 
$\mu$. $P$ is calculated up to first and second order in strong coupling 
contant, $\alpha_s$. Here we take $\Lambda=2.88\mu$ and zero quark mass 
approximation~[21].\\ \\ \\ \\ \\ \\}
\label{fig:Pvsmu}
\end{figure}

\begin{figure}[h]
\epsfig{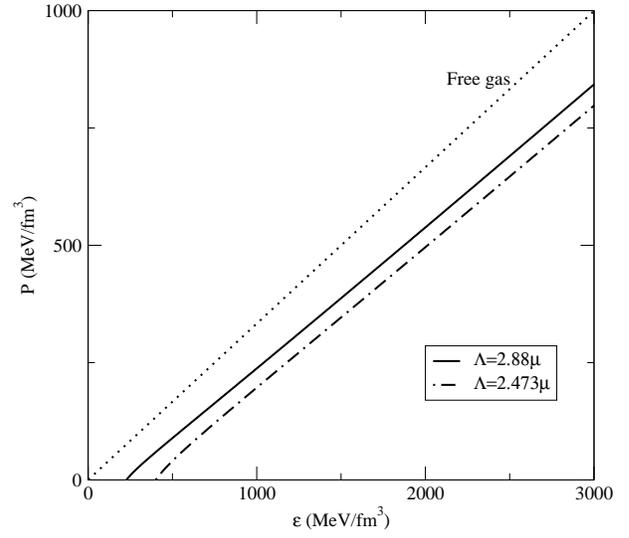}
\caption{Equation of state (pressure, $P$, vs energy density, $\epsilon$) 
for cold quark matter with massless quark approximation at various 
$\Lambda$, and free gas~[21].\\ \\ \\ \\ \\ \\}
\label{fig:PvsE}
\end{figure}

All the thermodynamic properties can be obtained from the thermodynamic 
potential if $\Lambda$ is fixed. It is believed that 
$\Lambda/\mu$ lies in the range between 2 and 3 \cite{Fraga}.
Fig.~\ref{fig:Pvsmu} shows the total pressure of strange quark matter 
relative to the pressure of an ideal gas, $P_0$, as a function of the
chemical potential $\mu$. Both the first and 
second order terms decrease the pressure of the strange quark matter 
relative to the ideal gas. It has been pointed out that using 
perturbation theory at 
zero pressure is invalid \cite{Fraga}. Numerical calculation shows 
that at zero pressure, $\alpha_s / \pi$ lies between 0.207 and 0.191, 
which is less than 1 
if $\Lambda/\mu$ lies between 2 and 3. 
The EOS's for $\Lambda/\mu=2.473, 2.88$ and free 
gas are shown in Fig.~\ref{fig:PvsE}.
It turns out that this EOS is very similar to the MIT Bag Model EOS,
with an effective Bag constant \cite{Fraga}, and we would have obtained
basically the same results using the latter.  None of our results in the 
cooling calculation depends on the validity of perturbative QCD in 
the compact star regime.

\section{STABILITY OF STRANGE QUARK MATTER}
\begin{figure}[h]
\epsfig{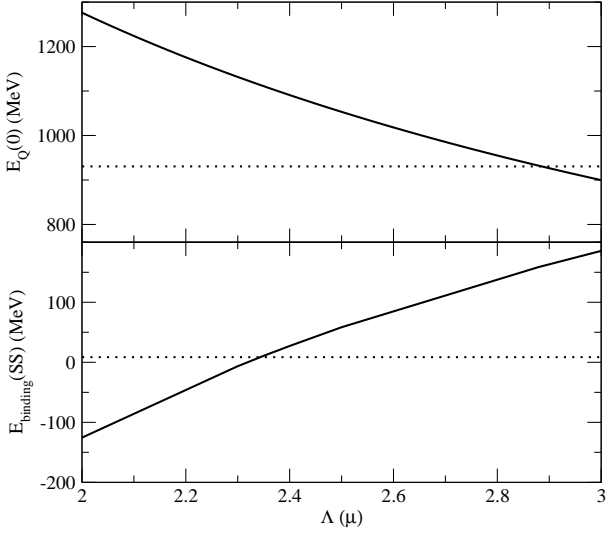}
\caption{Upper panel: Energy per baryon at zero pressure, 
${\mathcal{E_Q}}(0)$, as a function of 
$\Lambda/\mu$. The dotted line corresponds to the energy per baryon of 
$^{56}{\textrm{Fe}}$, which is equal to 930.4 MeV. Absolute stability of 
strange quark matter corresponds to $\Lambda \geq 2.880\mu$. Lower panel: 
Binding energy per baryon 
${\mathcal{E}}_{\textrm{\scriptsize{binding}}}($SS$)$ as a function of 
$\Lambda/\mu$. The dotted line corresponds to the binding energy of Fe, 
which is equal to 8.525 MeV. Global stability corresponds to $\Lambda/\mu 
\geq 2.35\mu$.
}
\label{fig:EvsLambda}
\end{figure}

A strong condition for strange quark matter to be the 
absolute ground state of cold matter is that the energy per baryon at 
zero pressure $\mathcal{E_Q}$(0) is less than that of $^{56}$Fe: 
\begin{equation}
{\mathcal{E_Q}}(0) < {\mathcal{E}}_{\textrm{Fe}} = 930.4{\textrm{ MeV}}.
\end{equation}
The upper panel of Fig.~\ref{fig:EvsLambda} shows that strange quark 
matter is absolutely 
stable for $\Lambda \geq 2.88 \mu$. At $\Lambda = 2.88 \mu$, the 
baryonic number density at zero pressure is about $1.52 n_0$, where $n_0$ 
is the normal nuclear matter number density. Therefore,  
strange quark matter can be the true ground state, and bare strange stars can 
exist if $\Lambda \geq 2.88 \mu$.

A weaker condition for strange quark matter to be stable  
is that the average energy per baryon of the bulk 
matter is less than that of $^{56}$Fe, ie., the binding energy per baryon of 
the bulk strange quark matter
${\mathcal{E}}_{\textrm{\scriptsize{binding}}}($SS$)$
is larger than that of $^{56}$Fe, 
${\mathcal{E}}_{\textrm{\scriptsize{binding}}}($Fe$)$: 
\begin{equation}
{\mathcal{E}}_{\textrm{\scriptsize{binding}}}(\textrm{SS})=
\frac{M_B-M_G}{A}>{\mathcal{E}}_{\textrm{\scriptsize{binding}}}(\textrm{Fe})
=8.525{\textrm{ MeV}}, 
\end{equation}
where $M_B$ and $M_G$ are the baryonic mass and gravitational mass of 
strange quark matter respectively, which are to be calculated by the 
Tolman-Oppenheimer-Volkov (TOV) equations. 
We have plotted the binding 
energy per baryon of the maximum mass star against $\Lambda$ in the lower panel of  
Fig.~\ref{fig:EvsLambda}. We found that the strange star can be globally 
stable compared to iron at infinity if $\Lambda > 2.35 \mu$.

\begin{figure}[h]
\epsfig{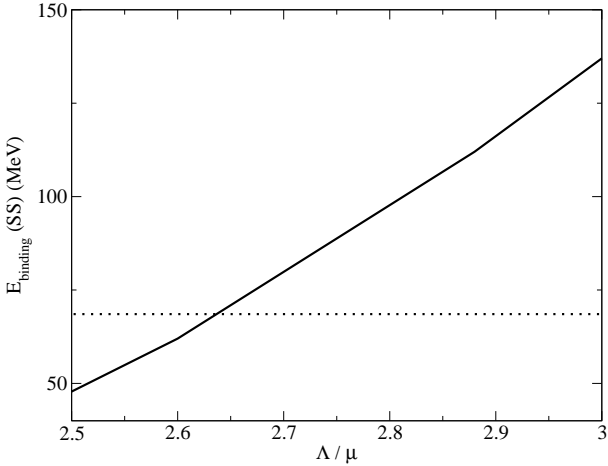}
\caption{
Binding energy per baryon
${\mathcal{E}}_{\textrm{\scriptsize{binding}}}($SS$)$ as a function of
$\Lambda/\mu$ for baryonic mass of strange stars being equal to that of a
$1.4 M_{\odot}$ neutron star with HV EOS.
The dotted line corresponds to the binding energy per baryon of the neutron
star, which is equal to 68 MeV. Global stability compared to the neutron
star corresponds to $\Lambda \geq 2.65 \mu$.}
\label{fig:stability:EvsLn}
\end{figure}

The weak stability condition above overestimates the stability of a 
strange star, since the gravitational binding of bulk $^{56}$Fe matter
is ignored.  A fairer condition is to compare the binding energy of
a strange star with that of a neutron star.  There are inevitably some
dependences on the EOS's used in such a comparison.  In 
Fig.~\ref{fig:stability:EvsLn} we show
the parameter regime in which strange stars are more stable than neutron
stars using one particular EOS (HV).  We have used several other EOS's
to reach a similar conclusion, that strange stars are more stable 
than neutron stars for $\Lambda \gtrsim 2.7 \mu$ 
[Table~\ref{table:stability:binding}].

\section{EFFECT OF MASSIVE STRANGE QUARKS}
In the density range of a quark star, i.e.~chemical potential $\sim 300-  
600$ MeV, the strange quark mass ($\sim 150$~MeV) may alter the EOS.  We consider 
the correction of the thermodynamic potential due to strange 
quark mass $m_s$ up to first order in 
$\alpha_s$. The masses of u and d quarks are small and can be neglected. 
The individual thermodynamic potentials are \cite{Glendenning}:
\begin{widetext}
\begin{eqnarray}
\Omega_f(\mu_f) & = & - \frac{\mu_f^4}{4 \pi^2} \left\{1-2\left( 
\frac{{\alpha}_{s}}{\pi}\right)-
\left[G+{N}_{f}\ln\left(\frac{{\alpha}_{s}}{\pi}\right)+
\left(11-\frac{2}{3}{N}_{f}\right)\ln
\frac{\Lambda}{\mu}\right]\left(\frac{{\alpha}_{s}}{\pi}\right)^2\right\}
\\
\Omega_s(\mu_s) & = & -\frac{1}{4\pi^2}\left\{
\mu_s(\mu_s^2-m_s^2)^{1/2}\left(\mu_s^2-\frac{5}{2}m_s^2\right)+ 
\frac{3}{2}m_s^4 \ln\left( 
\frac{\mu_s+(\mu_s^2-m_s^2)^{1/2}}{m_s}\right) \right.
\nonumber\\
                &   &
-2\left(\frac{\alpha_s}{\pi}\right)\left[  3 \left( 
\mu_s(\mu_s^2-m_s^2)^{1/2}-m_s^2 \ln\left( 
\frac{\mu_s+(\mu_s^2-m_s^2)^{1/2}}{m_s} \right) \right)^2
-2(\mu_s^2-m_s^2)^2
\right.
\nonumber\\
                &   &
\left. 
-3m_s^4 \ln^2\left( \frac{m_s}{\mu_s}\right) 
+ 6 \ln\left( \frac{\sigma}{\mu_s}\right)\left(\mu_s m_s^2 
(\mu_s^2-m_s^2)^{1/2}-m_s^4 \ln\left( 
\frac{\mu_s+(\mu_s^2-m_s^2)^{1/2}}{m_s}\right) \right) 
\right]
\nonumber\\
                &   &
\left.
-\mu^4\left[G+{N}_{f}\ln\left(\frac{{\alpha}_{s}}{\pi}\right)+
\left(11-\frac{2}{3}{N}_{f}\right)\ln
\frac{\Lambda}{\mu}\right]\left(\frac{{\alpha}_{s}}{\pi}\right)^2 
\right\},
\\
\Omega_e(\mu_e) & = & -\frac{\mu_e^4}{12\pi^2},
\end{eqnarray}
\end{widetext}
where $f=u$ or $d$, and $\sigma$ is the renormalization point for the 
strange quark mass. It 
is found that a suitable choice for $\sigma$ is $313$ MeV\cite{Farhi}. 

\begin{figure}[h]
\epsfig{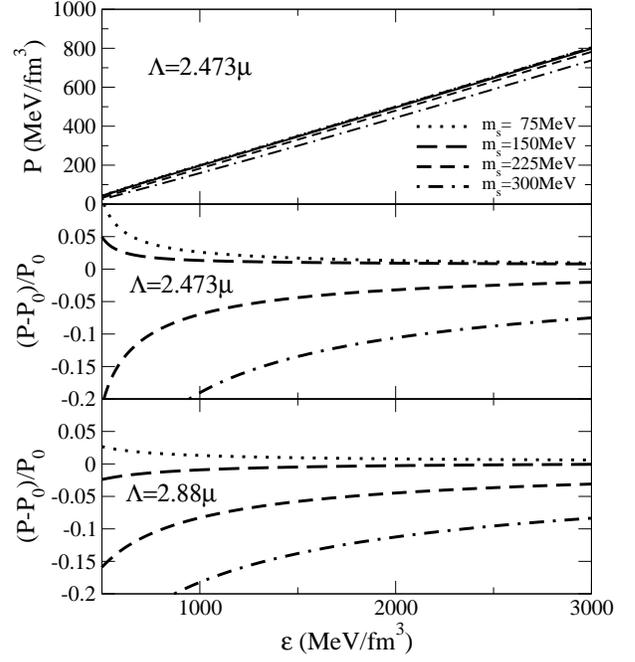}
\caption{Effect of strange quark mass, $m_s$, on the pressure $P$. 
$P_0$ is the zero quark mass pressure of strange quark matter about the 
same energy density $\epsilon$.} 
\label{fig:ms:PvsE}
\end{figure}

The EOS's are calculated for $\Lambda=2.473, 2.88$ and various $m_s$ 
numerically [Fig.~\ref{fig:ms:PvsE}]\cite{kawa-thesis}. 
For $m_s \leq 225$ MeV, the changes in pressure compared to massless EOS 
are less than $\sim5 \%$ in the high energy density limit. 
In the low energy density limit, the changes are significant because 
the strange quark mass is not small compared to the chemical potentials. 
Since the global structure of a compact star is mainly determined by the 
high density regime -- about several times the normal nuclear energy 
density -- of the EOS, the effect of the strange quark mass on the strange 
star structure is small. The correction to the maximum mass of quark 
stars due to a quark mass up to 150 (225) MeV for each series of $\Lambda$ 
is less than $3 (10)\%$  [Table \ref{table:ms:ms_corr}]. Since the 
experimental data shows that the mass of 
strange quark is $\sim 150$~MeV, we can safely ignore the quark mass when 
calculating the global properties of a quark star.

A recent study shows that quark matter in the Color-Flavor Locked (CFL) 
phase can be electrically neutral, even though the quark masses are unequal. 
Electrons are not needed to maintain the charge neutrality 
\cite{Rajagopal}. 
However, the CFL phase occurs at the very high density regime. Electrons 
or positrons may still be present near the surface of a quark star, 
and they are important for its cooling behaviour.
More detailed discussion of the effect of massive strange quarks and 
its physical implications can be found in 
Ref.~\cite{Farhi,kawa-thesis,Peng,Madsen}.

\section{PHASE TRANSITION FROM STRANGE STARS TO NEUTRON STARS}
It has long been suggested that strange stars can be formed from a
phase transition of neutron stars to strange stars due to an abrupt
increase in density \cite{cheng1,cheng2}. However, from the theoretical
point of
view, formation of quark-gluon plasma is favoured when both temperature
and chemical potential are high enough \cite{shuryak}. Hence it is
reasonable to suggest that strange stars are formed in supernovae where
both high temperature and density are achieved \cite{kawa1}. If the strange 
quark
matter is absolutely stable for high density and zero temperature, i.e.,
$\Lambda/\mu \geq 2.88$, the quark star will remain in the quark phase
even when it cools down. If the strange star is energetically less
stable than a neutron star below some temperature, it will cool down to the
phase transition temperature, $T_p$, and change into a neutron star
containing
ordinary baryons. If the baryonic mass $M_{B}$ is conserved during the
phase transition, the total conversion energy
$E_{\textrm{\scriptsize{conv}}}$ released is given by:
\begin{equation}
E_{\textrm{\scriptsize{conv}}}=[M_{G}({\rm SS})-M_{G}({\rm NS})]c^2,
\end{equation}
where $M_{G}({\rm SS})$ and $M_{G}({\rm NS})$ are the gravitational
masses of the strange star and neutron star respectively \cite{Bombaci}.
Whether a phase transition can occur and how much energy is released
depend on both the EOS's of quark matter and nuclear matter. We are
interested in the possibility that strange quark matter is only stable
for $T > T_p$, and so we choose a $\Lambda/\mu < 2.7$, so that when the hot
strange star cools to low temperature, it will convert to a neutron star.
For $\Lambda/\mu = 2.473$, the maximum gravitational mass is $1.516
M_{\odot}$ with a baryonic mass of $1.60 M_{\odot}$ and radius 8.54 km.
We will use this set of parameters in the calculation of the cooling
behaviour because the maximum mass is close to observational data of
compact stars. In fact, we have used other values of $\Lambda/\mu$, and
the cooling behaviour is qualitatively similar, as long as the star
undergoes a phase transition. The cold EOS of the neutron stars will
only determine the conversion energy. The cooling behaviour is determined
by the cooling mechanisms and heat capacities in our calculations. We
simply choose the HFV EOS based on the relativistic
Hartree-Fock approximation \cite{Weber} to calculate the conversion
energy. A typical neutron star with
$M_{B}=1.60M_{\odot}$ and $M_{G}=1.40M_{\odot}$ is chosen \cite{Weber} in
the cooling calculations.
For comparision, the energy released for different $\Lambda/\mu$, together
with several commonly used neutron star EOS's are summarized in Table
\ref{table:cooling:conv}.
For $\Lambda/\mu=2.473$, typically $10^{53}$ erg is released during the
process, which depends only weakly on the nuclear matter EOS. We propose 
that it can be an energy source of Gamma Ray Bursts 
(GRB's)~\cite{kawa-thesis}.

\section{COOLING PROPERTIES}
It is believed that in an ultrarelativistic heavy-ion collision, a hot
quark-gluon plasma is formed initially, which then cools down to
the phase transition temperature $T_p$ and goes into the
mixed phase, if the QCD phase transition is first order. When all the quark
matter has hadronized, the temperature drops again \cite{Srivastava}.
We borrow this idea to describe the
cooling of a strange star. When the strange star is born, the
temperature can be as high as a few times $10^{11}$ K \cite{phillips}.
The surface is so
hot that all the materials, other than strange-quark matter, are
evaporated leaving the strange star nearly bare without any crust
\cite{Usov2001}. Since the thermal conductivity of strange quark matter
is very high and the density profile of the strange star is very flat, we 
will
take the uniform temperature and density approximation.
Hence the strange star cools down according to the equation:
\begin{equation}
C_q\frac{dT}{dt}=-L_q ,
\end{equation}
where $C_q$ is the total heat capacity of all the species in quark
matter, and $L_q$ is the total luminosity of the star.
When the temperature drops to $T_p$, the strange star undergoes a phase
transition and a latent heat $E_{\textrm{\scriptsize{conv}}}$ is released.
We simply take $T_p$ to be constant in density throughout the star, which 
has a much smaller 
value than the zero chemical potential value of about 150 MeV  because
of the high chemical potential.  During the
phase transition, we assume that the quark and neutron matter are distributed
uniformly and calculate the luminosity of the mixed phase by the
weighted average of that of the quark matter and the neutron matter
[Section~\ref{section:PT}].
When the strange star has lost all its latent heat and converted completely
to a neutron star, it then follows the standard
cooling of a neutron star with an initial temperature of $T_p$.
The detailed thermal evolution is governed by
several energy transport equations. We adopt a simple model that a
neutron star has a uniform temperature core with high conductivity and
two layers of crust, the inner crust and the outer crust, which transport
heat not as effectively as the core or quark matter. The typical thickness
of the crust is $\sim 10\%$ of the radius, and we can use the parallel-plane
approximation to describe the thermal evolution of the inner crust. The
thermal history of the inner crust can be described by a heat conduction
equation:
\begin{equation}
c_{\textrm{\scriptsize{crust}}} \frac{\partial T}{\partial t} = 
\frac{\partial}{\partial r}
\left( K\frac{\partial T}{\partial r} \right) - \epsilon_{\nu},
\end{equation}
where $c_{\textrm{\scriptsize{crust}}}$ is the specific heat of the inner
crust, $K$ is the effective thermal conductivity, and $\epsilon_{\nu}$ is
the neutrino emissivity. As a rule of thumb, the effective surface
temperature $T_s$ and the temperature at the interface of inner and outer
crust, $T_b$, are related by \cite{Gudmundsson}:
\begin{equation}
T_{b8}=1.288(T_{s6}^4/g_{s14})^{0.455},
\end{equation}
where $g_{s14}$ is the surface gravity in the unit of $10^{14}$ cm
$\textrm{s}^{-2}$, $T_{b8}$ is the temperature between the inner and
outer crusts in the unit of $10^8$ K, and $T_{s6}$ is the effective surface
temperature in the unit of $10^6$ K. The luminosity at the stellar surface,
$L_{\textrm{\scriptsize{surface}}}$, is equal to the heat flux at the
interface of inner and outer crusts:
\begin{equation}
-K\frac{\partial T}{\partial
r}=L_{\textrm{\scriptsize{surface}}}/(4\pi R^2),
\end{equation}
where $R$ is the radius of the star. The boundary condition at the
interface of the core and inner crust is:
\begin{equation}
C_{\textrm{\scriptsize{core}}}\frac{\partial T}{\partial
t}=-K\frac{\partial T}{\partial
r}A_{\textrm{\scriptsize{core}}}-L_{\nu}^{\textrm{\scriptsize{core}}},
\end{equation}
where $C_{\textrm{\scriptsize{core}}}$ is the total heat capacity of the
core,
$A_{\textrm{\scriptsize{core}}}$ is
the surface area of the core, and $L_{\nu}^{\textrm{\scriptsize{core}}}$ is
the total neutrino luminosity of the core.

\subsection{Heat capacity of quark stars}
The total heat capacity is the sum of the heat capacities of all
species in the star. Without the effect of superfluidity, the quark
matter can be considered as a free Fermi gas, and the specific heat of
quark matter is given by \cite{Iwamoto,rebhan}:
\begin{equation}
{c}_{q}=2.5\times 10^{20}(\rho / {\rho}_{0})^{2/3}{T}_{9}
\textrm{ erg cm}^{-3}\textrm{ K}^{-1},
\end{equation}
where $\rho$ is the baryon density and $\rho_0=0.17 \textrm{ fm}^{-3}$ is
the nuclear matter density. Some authors may use energy density instead
of number density. The difference is only of a numerical factor and does
not change the cooling curve much. In the superfluid state, the quarks
will form Cooper pairs. The specific heat will be modified as
\cite{Horvath,Maxwell}:
\begin{equation}
c_{q}^{\textrm{\scriptsize{sf}}}=\left\{
\begin{array}{ll}
\frac{3.15c_{q}}{{\tilde T}}
e^{-\frac{1.76}{\tilde T}} \left[2.5-1.66 {\tilde T}
+3.64 {\tilde T}^2 \right] & \textrm{for $0.2\leq {\tilde T}\leq 1$}\\
0      & \textrm{for ${\tilde T}<0.2$,}
\end{array}
\right.
\end{equation}
where $k_BT_c=\Delta /1.76$, ${\tilde T}=T/T_c$ and $\Delta$ is the energy
gap in MeV in BCS
theory. It has been argued that for quark matter, even with unequal quark
masses, in the Color-Flavor Locked (CFL) phase in which all the three
flavors and colors are paired, quark matter is automatically charge neutral
and no electrons are required \cite{Rajagopal}. However, for sufficiently
large strange quark mass and the
relatively low density regime near the stellar surface, the 2 color-flavor
SuperConductor (2SC) phase is expected to be preferred. Therefore in a real
strange star, electrons are believed to be present. The contribution of
electrons can be parametrized by the electron fraction $Y_e$, which
depends on the model of strange stars. We choose $Y_e=0.001$ as a typical
value. The specific heat capacity of electrons in the strange star phase
is given by \cite{Ng}:
\begin{equation}
c_{e}=1.7\times 10^{20} \left(\frac{Y_{e}\rho}{\rho_{0}} \right)^{2/3}T_{9}
\textrm{ erg cm}^{-3}\textrm{ K}^{-1}.
\end{equation}
The heat capacity of electrons is unaffected by the superfluidity of quark
matter. Hence it dominates the total heat capacity of the strange star
when the temperature drops below $T_c$.

\subsection{Luminosity of quark stars}
\begin{figure}[h]
\epsfig{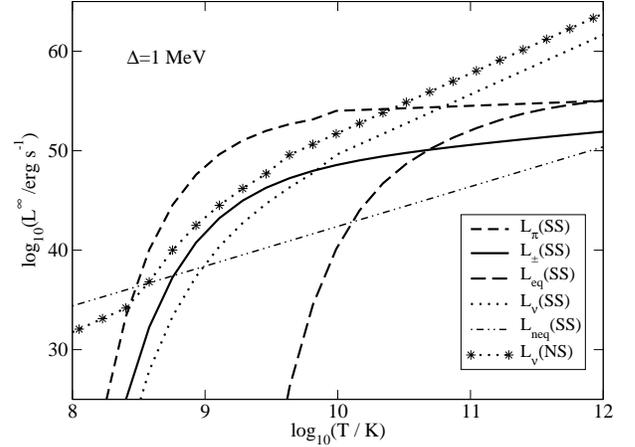}
\caption{\small Observed luminosity $L^{\infty}$ as a function of
temperature
$T$ for small superfluid gap model in the quark phase $\Delta=1$~MeV.
$L_{\pi}, L_{\pm}, L_{eq}, L_{\nu}$ and $L_{neq}$ denote the pion,
$e^+e^-$, thermal equilibrium radiation, neutrino emission and
non-equilibrium blackbody radiation luminosity respectively. For
comparison, the neutrino luminosity of neutron star phase is also
plotted. SS and NS correspond to strange
stars and neutron stars respectively.~[8]
}
\label{fig:cooling:lum1}
\end{figure}

\begin{figure}[h]
\epsfig{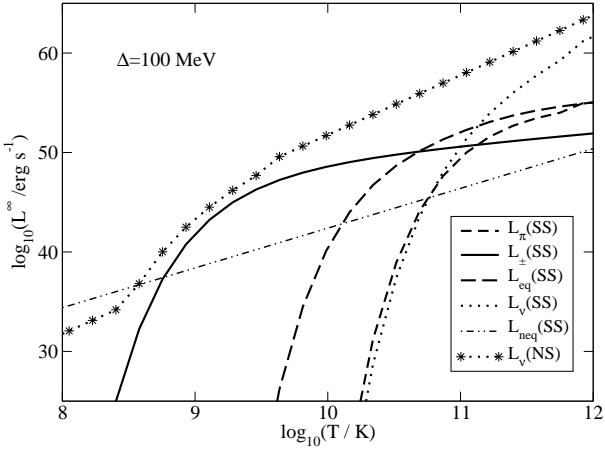}
\caption{\small Same as Fig.~\ref{fig:cooling:lum1}, but for a large
superfluid gap model in the quark phase $\Delta=100$~MeV.~[8]
}
\label{fig:cooling:lum2}
\end{figure}

The total luminosity is the sum of contributions from all energy
emission mechanisms,
including both photon and neutrino emmission. The cooling of quark-gluon
plasma has been studied for many years \cite{Weber} and can be divided into
fast and slow cooling processes. Two of the most popular fast cooling
processes are the  electron-positron pair production and quark URCA
process. Two well known slow cooling processes are thermal
equilibrium and non-equilibrium blackbody radiation. We have included all
these cooling processes in the our calculations.
The recently proposed pion production in a strange star is found
to be a very effective cooling mechanism, and it may explain the energy
supply of Gamma Ray Bursts (GRB's) \cite{Ng}. Therefore we also studied the 
effect of
the pion production cooling. Luminosities of various cooling mechanisms
are plotted in Fig.~\ref{fig:cooling:lum1} and \ref{fig:cooling:lum2}.

\subsubsection{Pion Production}

It is known that for a hot
quark-gluon plasma described by a Cloudy Bag Model, pions can be produced
through two mechanisms: thermal excitation and collisions between quarks
and the bag surface \cite{Ng,Csernai,Muller}. When the pions
leave the quark star surface, they will decay into photons and $e^+e^-$
pairs through the following channels:
\begin{equation}
\pi^0\rightarrow 2\gamma \leftrightarrow e^+e^-,
\end{equation}
\begin{equation}
\pi^{\pm}\rightarrow \mu^{\pm}+\nu_{\mu},
\end{equation}
\begin{equation}
\mu^{\pm}\rightarrow e^{\pm}+\nu_e+\nu_{\mu}.
\end{equation}
It has been shown that the production of pions is a very powerful source
of $e^+e^-$ pairs and photons, with luminosity $\sim 10^{54} \textrm{ erg
s}^{-1}$. Such a powerful source of $e^+e^-$ pairs and photons may
serve as the source engine of $\gamma$-ray bursts. The pion emissivity is
estimated to be \cite{Ng}:
\begin{equation}
L_{\pi}=\rho_{\pi}\sqrt{\frac{2k_B T}{m_{\pi}}}4\pi R^2,
\end{equation}
where $\rho_{\pi}=7.1\times 10^{31} \textrm{ erg cm}^{-3}$ is the energy
density of pion field at the stellar surface, $T$ is the temperature of
the quark star, which is taken to be uniform, and $m_{\pi}=140$ MeV is the
pion mass. In the superfluid state, the collisions between quarks are
suppressed due to the pair up of quarks, and the pion luminosity is then
reduced by a factor of $\exp (-\Delta /T)$.

\subsubsection{Quark URCA Process}

The dominating neutrino emission process is the quark URCA process:
\begin{equation}
d\rightarrow u+e+\overline{\nu}_e,
\end{equation}
\begin{equation}
u+e\rightarrow d+\nu_e.
\end{equation}
The neutrino emissivity was estimated as \cite{Iwamoto}:
\begin{equation}
\epsilon_{d}\simeq 2.2\times
10^{26}\alpha_{s} \left(\frac{\rho}{\rho_{o}} \right)Y_{e}^{1/3}T_{9}^{6}
\textrm{ erg cm}^{-3}\textrm{s}^{-1},
\label{eq:cooling:qURCA}
\end{equation}
where $\alpha_s$ is the strong coupling constant, and we have chosen
$\alpha_s=0.4$ as a constant value throughout the quark star. Note that
the definition of the strong coupling constant is different from that of
Iwamoto's.  In the superfluid state, the neutrino emissivity is
suppressed by a factor of $\exp(-\Delta /T)$.

\subsubsection{Electron-positron Pair Production}

It has been pointed out that the bare surface of a hot strange star is a
powerful source of $e^+e^-$ pairs due to the strong electric field at the
surface \cite{Usov1998}.
The $e^+e^-$ pair production rate is \cite{Usov2001}:
\begin{equation}
\dot{n}_{\pm} \simeq 10^{39} T_9^3
\exp(-11.9/T_9)
J(\xi) {\textrm s}^{-1},
\end{equation}
where
\begin{equation}
J(\xi)=\frac{1}{3}\frac{\xi^3
\ln(1+2\xi^{-1})}{(1+0.074\xi)^3}+\frac{\pi^5}{6}\frac{\xi^4}{(13.9+\xi)^4},
\end{equation}
\begin{equation}
\xi=2\sqrt{\frac{\alpha}{\pi}}\frac{\varepsilon_F}{k_B T} \simeq 0.1
\frac{\varepsilon}{k_B T},
\end{equation}
$\alpha=1/137$ is the fine-structure constant and $\varepsilon_F=18$ MeV
is the Fermi energy of electrons. The luminosity of $e^+e^-$ pairs is
given by:
\begin{equation}
F_{\pm}\simeq \varepsilon_{\pm}\dot{n}_{\pm},
\end{equation}
where $\varepsilon_{\pm} \simeq m_e c^2+k_B T$ is the mean energy
of created particles.

\subsubsection{Thermal Equilibrium Radiation}

The thermal equilibrium radiation of frequency $\omega$ less than the plasma
frequency $\omega_p \simeq 20-25$ MeV is greatly suppressed due to the
very high density of the quark-gluon plasma\cite{Alcock,Usov2001}.
The luminosity of thermal equilibrium photons is\cite{Usov2001}:
\begin{equation}
F_{\textrm{\scriptsize{eq}}}=\int_{\omega_p}^{\infty} d\omega
\frac{\omega(\omega^2-\omega_p^2)g(\omega)}{\exp (\omega/T)-1},
\label{eq:cooling:eq_radiation}
\end{equation}
where
\begin{equation}
g(\omega)=\frac{1}{2\pi}\int_0^{\pi/2}d\theta \sin \theta \cos \theta
D(\omega,\theta),
\end{equation}
$D(\omega,\theta)=1-(R_{\perp}+R_{\parallel})$ is the coefficient of
radiation transmission from the quark-gluon plasma to vacuum, with
\begin{equation}
R_{\perp}=\frac{\sin^2 (\theta-\theta_0)}{\sin^2(\theta+\theta_0)},
R_{\parallel}=\frac{\tan^2 (\theta-\theta_0)}{\tan^2(\theta+\theta_0)},
\end{equation}
\begin{equation}
\theta_0\equiv\arcsin \left[ \sin \theta \sqrt{1-(\omega_p/\omega)^2}
\right].
\end{equation}

\subsubsection{Non-equilibrium Blackbody Radiation}

The above processes are very powerful sources of energy emission and will
dominate the cooling process at very high
temperature. Once the temperature drops, the cooling process will be
dominated by the relatively low power non-equilibrium blackbody radiation
\cite{cheng2003}:
\begin{equation}
L_{\textrm{\scriptsize{neq}}}\approx 10^{-6}L_{bb},
\end{equation}
where $L_{bb}=4\pi R^{2}\sigma T^{4}$ is the blackbody radiation luminosity.

\subsection{Microphysics of the neutron star cooling}

There are many different models of neutron star cooling. Since we mainly
focus on the cooling of the quark phase and examine the phase transition
process qualitatively, the cooling of neutron stars can be taken from
any model available in the literature. We simply adopt the model of
neutron stars described by Ng \cite{Ng,Maxwell}. The heat capacities
of neutron matter in both normal and superfluid states, and of the
electrons are:
\begin{equation}
c_n = 2.3\times 10^{39}M_* \rho_{14}^{-2/3} T_9 \textrm{ erg K}^{-1}
\textrm{~~~for ${\tilde T} > 1$,}
\end{equation}
\begin{equation}
c_{n}^{\textrm{\scriptsize{sf}}} = \frac{3.15c_{n}}{\tilde T}
e^{-\frac{1.76}{\tilde T}} \left[2.5-1.66 {\tilde T}
+3.64 {\tilde T}^2 \right]
{\rm ~~~for} ~~0.2 \le {\tilde T} \le 1
\end{equation}
and
\begin{equation}
c_e=1.9\times 10^{37}M_* \rho_{14}^{1/3} T_9 \textrm{ erg K}^{-1},
\end{equation}
where $M_*$ is the mass of neutron star in units of solar mass,
$\rho_{14}=\rho/10^{14}\textrm{ g cm}^{-3}$, and we take $T_c=3.2 \times
10^{9}$ K for our calculations.

The neutrino emission mechanisms are the direct URCA process:
\begin{equation}
n \rightarrow p + e^- + \overline{\nu}_e,
\end{equation}
\begin{equation}
p + e^- \rightarrow n + \nu_e,
\end{equation}
with the neutrino emissivity \cite{Ng,Lattimer}:
\begin{equation}
\epsilon_{\textrm{{\scriptsize URCA}}} = 4.00\times 10^{27} (Y_e
\rho/\rho_0)^{1/3}T_9^6 \textrm{ erg cm}^{-3}\textrm{s}^{-1}
\textrm{ for }T>T_c,
\label{eq:cooling:dURCA}
\end{equation}
\begin{equation}
\epsilon_{\textrm{{\scriptsize URCA}}}^{\textrm{\scriptsize{sf}}} =
\epsilon_{\textrm{{\scriptsize URCA}}} \exp(-\Delta/T) \textrm{ for }T<T_c,
\end{equation}
with $Y_e=0.1$, electron-proton Coulomb scattering in the crust:
\begin{equation}
e + p \rightarrow e + p + \nu + \overline{\nu}
\end{equation}
with luminosity \cite{Ng,Festa??}:
\begin{equation}
L_{\nu}^{\textrm{{\scriptsize cr}}} = 1.7\times 10^{39}M_*
(M_{\textrm{\scriptsize{cr}}}/M) T_9^6 \textrm{ erg s}^{-1},
\label{eq:cooling:crust}
\end{equation}
where $(M_{\textrm{{\scriptsize cr}}}/M)$ is the fractional mass of the
crust $\approx 5\%$, and the neutrino bremsstrahlung process:
\begin{equation}
n+n \rightarrow n+n+\nu+\overline{\nu}
\end{equation}
with luminosity \cite{Ng}:
\begin{eqnarray}
L_{\nu}^{\textrm{\scriptsize{nn}}} = & 4.3\times
10^{38}\rho_{14}^{1/3}T_9^8 \textrm{ erg s}^{-1} & \textrm{ for } T > T_c
\nonumber\\
                                   = & 0
                                                 & \textrm{ for } T \leq T_c.
\label{eq:cooling:bras}
\end{eqnarray}

The surface luminosity will be of the blackbody radiation with the
effective surface temperature $T_s$:
\begin{equation}
L_{bb}=4\pi R^{2}\sigma T_s^{4}.
\end{equation}
The blackbody radiation will be the dominating cooling mechanism after
the neutrino emission processes are switched off.

For the thermal conductivity of the inner crust, we simply choose a
temperature dependent model \cite{Lindblom}:
\begin{equation}
K=\frac{2.8 \times 10^{20}}{T_{10}} \textrm{ erg cm}^{-1}\textrm{ 
s}^{-1}\textrm{ K}^{-1}. 
\end{equation}
The choice of thermal conductivity will not be important after the epoch
of thermal relaxation, which is of the order $10-100$ years. The
temperature of the inner crust and the core will be uniform after that.

\subsection{Handling of the Phase Transition}
\label{section:PT}

The actual situation during the phase transition is very complicated,
involving detailed hydrodynamical simulations, how quark matter is
transformed into ordinary hadronic matter and so on. Many authors have
discussed the theoretical modeling of the phase transition in heavy-ion
collisions, neutron star-strange star phase transition and the
cosmological quark-hadron transition, but the uncertainties are of course
very large at this stage. Without involving the details, we simply
assume that the phase transition is first order and
evolves quasi-statically. The quark and neutron matter are distributed
uniformly in the mixed phase. We develop a simple model to describe the
luminosities of the mixed phase in order to capture the main features just
before and after the phase transition. The detailed results during the
phase transition is of course inaccurate, but the main features should be
approximately correct, as long as the phase transition does not disrupt the
neutron star completely.

The energy loss during the phase transition is:
\begin{equation}
E_{\rm loss}=\int_{t_0^{\rm PT}}^{t} L_q^{\rm PT}+L_n^{\rm PT} ~dt,
\end{equation}
where $L_q^{\rm PT}$ and $L_n^{\rm PT}$ are the total
luminosity of
the quark matter contribution and nuclear matter contribution during the
phase transition respectively, $t_0^{\rm PT}$ is the starting time of the
phase transition.
We define the fractions of total baryonic number of quarks and neutrons in
the mixed phase to be $A_{\rm SS}$ and $A_{\rm NS}$ respectively. $A_{\rm
SS}$ and $A_{\rm NS}$ will be related to the conversion energy $E_{\rm
conv}$ and energy loss during the phase transition $E_{\rm loss}$ by:
\begin{equation}
A_{\rm SS} = \frac{E_{\rm conv}-E_{\rm loss}}{E_{\rm conv}}
\end{equation}
and
\begin{equation}
A_{\rm NS} = \frac{E_{\rm loss}}{E_{\rm conv}}.
\end{equation}

\subsubsection{Pion Production, $e^+e^-$ Pair Production}
Since the pion production and $e^+e^-$ pair production are the features
of a hot quark star, we expect the production rates to decrease with
the fraction of quark number during the phase transition. We simply treat
the luminosities of both contributions to be:
\begin{equation}
L_{\pi (e^+e^-)}^{\rm PT}= A_{\rm SS} L_{\pi (e^+e^-)}.
\end{equation}

\subsubsection{Thermal Equilibrium Radiation}
The plasma frequency is related to the baryon number density as
\cite{Usov2001}:
\begin{equation}
\omega_p=\left(\frac{8\pi}{3}\frac{e^2 c^2 n_b^2}{\mu}\right)^{1/2},
\end{equation}
where $\mu \simeq \hbar c (\pi^2 n_b)^{1/3}$ is the chemical potential.
Hence:
\begin{equation}
\omega_p \propto n_b^{1/3}.
\end{equation}
For a typical strange star, $\omega_p \simeq 20$ MeV. Therefore during the
phase transition, the plasma frequency will be related to the fraction of
quark number as:
\begin{equation}
\omega_p^{\rm PT} \simeq 20 {A_{\rm SS}}^{1/3}  {\rm ~MeV}.
\end{equation}
The luminosity of thermal equilibrium photons during the
phase transition is given by Eq.(\ref{eq:cooling:eq_radiation}) where 
$\omega_p=\omega_p^{\rm PT}$.
We can see that the lower limit becomes smaller during the phase
transition. The luminosity reduces to that of thermal blackbody radiation
at the end of the phase transition. Hence we expect an increase of
equilibrium photon luminosity as a signature of the phase transition.

\subsubsection{Quark URCA Process}
From Eq.(\ref{eq:cooling:qURCA}), the neutrino emissivity of a quark star
$\epsilon_d \propto \rho$. Hence,
\begin{equation}
\epsilon_d^{\rm PT}=\epsilon_{d} A_{\rm SS}.
\end{equation}

\subsubsection{Neutrino Emission Process of the Neutron Matter}
For the neutrino emissivity of neutron matter, we use the
fractional density of neutron matter in the
density dependent terms of Eq.(\ref{eq:cooling:dURCA}) and
(\ref{eq:cooling:bras}), i.e.:
\begin{equation}
\rho \rightarrow A_{\rm NS} \rho.
\end{equation}
For the electron-proton Coulomb scattering of
Eq.(\ref{eq:cooling:crust}), we simply calculate its contribution with a
weighted mean:
\begin{equation}
L_{\nu}^{\rm cr ~PT}=L_{\nu}^{\rm cr} A_{\rm NS}.
\end{equation}
Lattimer $et$ $al.$ have pointed out that direct URCA process can only
be switched on in a neutron star at sufficiently high central density
\cite{Lattimer}. For a wide range of parameters, the required central
density is of the order of nuclear matter density. We switch on the
direct URCA process only when the average density of neutron matter is
above the nuclear matter density. 

\subsubsection{Non-equilibrium Blackbody Radiation}
Since the non-equilibrium blackbody radiation is not the
dominating cooling mechanism in the temperature range of 
$10^9-10^{11}$~K, it can be neglected in the mixed phase.

\section{The models}
When a strange star is born just after the stellar collapse, its 
temperature is very high $\sim 10^{11}$ K $\sim 40$ MeV~\cite{supernova}. We 
choose an initial temperature $T_i=40$~MeV. The EOS of 
perturbative QCD with massless quarks and temperature correction up to 
first order in $\alpha_s$ is \cite{Gentile}: 
\begin{eqnarray}
P & = & \frac{8\pi^2}{45}T^4 \left(1-\frac{15}{4}\frac{\alpha_s}{\pi}\right)
+\sum \left[\frac{7}{60}\pi^2 T^4 \left( 1- 
\frac{50}{21}\frac{\alpha_s}{\pi} \right) \right. \nonumber\\
 & & \left. +\left(\frac{1}{2}T^2 
\mu_f^2+\frac{1}{4\pi}\mu_f^4 \right) \left( 1-2\frac{\alpha_s}{\pi} \right) 
\right].
\end{eqnarray}
Since just after the collapse, $T \sim 40$ MeV $\ll \mu \sim 300$ MeV at the 
surface, the zero temperature EOS is appropriate for calculating the 
structure of the star. In our cooling model, the only free parameter of 
the EOS in the quark phase is $\Lambda$. It will determine whether a 
phase transition can occur and how much latent heat can be released. For a 
$1.4M_{\odot}$ compact star, the cooling curves are mainly affected 
by the cooling mechanisms and heat capacities of the material, rather 
than $\Lambda$, which mainly affects the structure of the compact star. 
Therefore we only choose one value of this parameter, $\Lambda = 2.473 \mu$, 
for the cooling calculations. The phase transition temperature $T_p$ 
involves detailed study of the baryon phase diagram, which is still 
highly uncertain. 
The latest results \cite{fodor,allton} indicate that $T_p$ drops from its 
zero 
density value of 140 MeV to a few MeV for density of a few times nuclear 
matter density. Hence we treat $T_p$ as a parameter and study the 
cooling behaviours for different $T_p$. 
We assume a constant $T_p$ in density throughout the star and present 
results 
for $T_p=1,10$~MeV for comparision. The remaining ingredients are the 
heat capacities and cooling mechanisms of the star. For the quark phase, 
the URCA process, thermal equilibrium and non-equilibrium radiation are 
believed to occur for a bare strange star. We include all these processes 
in our calculations, and we choose other 
combinations of cooling processes to investigate the effect of each cooling 
mechanism. For the superfluid phase of quark matter, we choose 
$\Delta = 1$ MeV for a small gap model and $\Delta = 100$ MeV for a large 
gap model. The various models presented in this paper are summarized in 
Table~\ref{model}.

\section{Results}
The strange star luminosity $L$ discussed above is
typically many orders of magnitude higher than $10^{37}$ erg/s
and may be as high as $10^{53}$ erg/s. In this case, the
outflowing wind is optically thick, and at $L> 10^{42}-10^{43}$
erg/s the spectrum of emergent photons is nearly a blackbody
spectrum \cite{aksenov1,aksenov2}. Therefore, for a new born strange star, 
the pion production, equilibrium radiation, and electron-positron 
pair production cannot be distinguished observationally in spite 
of the fact that the characteristic gamma-ray energies at these processes
near the strange star surface differ significantly. 
They can be distinguished observationally only if the phase transition 
temperature is very low, such that the strange star can exist in a 
relatively low temperature. 
The observables are the luminosity and the 
surface temperature at infinity, $L^{\infty}$ and 
$T_s^{\infty}$, which are related to the stellar surface values, $L$ and 
$T_s$ \cite{Tsuruta1998}: 
\begin{equation}
T_s^{\infty}=e^{\phi_s}T_s,
\end{equation}
\begin{equation}
L^{\infty}=e^{2\phi_s}L,
\end{equation}
where $e^{\phi_s}=\sqrt{1-2M/R}$ is the gravitational redshift at the 
stellar surface. The various cooling curves for different models are 
shown in Fig.~8.1 - 8.4. Models a to c are small superfluid gap 
models, 
while Models d to f are large superfluid gap models.
For small (large) superfluidity gap models, the phase transition 
temperature considered here is well above (below) the superfluid gap. We 
can get some insight of the effect of superfluidity on the 
signatures of phase transition.

\subsection{Model a, b}

The cooling curves for Model a and Model b are nearly the same. It is
because both of them have the same thermal properties, and all the cooling
mechanisms are switched on, except for Model a that the $e^-e^+$ pair
production is not considered. Since for the small superfluid gap
$\Delta=1$ MeV, pion emission is not suppressed for $T>T_c$. The pion
emissivity dominates the $e^-e^+$ pair production by several orders of
magnitude at high temperature. At low temperature, the cooling process
will be dominated by the blackbody radiation. Hence both models will be
dominantly cooled by the same mechanisms and have the same cooling
history. Only if the phase transition temperature is very low, we can 
distinguish them by studying the spectrum, in
which Model b shows the character of $e^-e^+$
production. In these models, for $T_p=1$ MeV,  the photon luminosity can
maintain up to $10^{50}-10^{54}$ erg s$^{-1}$ for about 1 s when the
star is in the quark and mixed phases. Such a `long'
duration of extremely high luminosity, compared to Ng $et$ $al.$'s model
which gives a luminosity that drops below
$10^{50}$ erg s$^{-1}$
within 0.01 s in the extreme case, is dominated by pion emission and
maintained by the latent heat of the star. This violent fireball easily
supplies the energy required for GRB's. The neutrino
luminosity drops
from $10^{57}$ to $10^{48}$ erg s$^{-1}$ within $0.001$ s and then rises
to $10^{51}$ erg s$^{-1}$ during the mixed phase which lasts for $\sim 1$ s.

This scenario of a second burst of neutrinos can be compared to two
previous similar proposals \cite{Benvenuto,Aguilera}.
In our model,
the burst is due to the phase transition from a quark star to a neutron star,
which has a higher neutrino emissivity, whereas in previous proposals, the
second burst accompanies the phase transition from a neutron star to a quark
star.  In Benvenuto's theory, the phase transition is delayed by a few seconds
after the core bounce due to the presence of the neutrinos \cite{Benvenuto}.
In Aguilera $et$ $al.$'s theory \cite{Aguilera},
the burst is due to the initial trapping of neutrinos when the
temperature is high and their sudden release when the quark star cools.
If quark matter is not as stable as nuclear matter at low temperature, then
there should be yet another phase transition back to nuclear matter, which is
what we focus on, and the ``second'' neutrino burst we propose is then the
``third'' neutrino burst.

If multiple neutrino bursts are observed, as may indeed be the case for
the Kamiokande data for SN1987A \cite{Hirata}, whether the compact star changes
from the quark phase to neutron phase (our model) or the other way around
can be distinguished observationally in at least two ways.
First, our model predicts that the cooling is much faster {\it before}
the phase transition, but it will become slower {\it after} it.
Second, the size of the post-phase-transition
compact star, being a normal neutron star, would be larger in our model.

For $T_p=10$ MeV, the high photon luminosity can only be maintained for $\sim
0.02$ s, since the neutrino luminosity is very high during the phase
transition, in which it releases the latent heat within a short time. The
rises in neutrino flux are about two orders of magnitudes, but the time
scale is so short that they are likely to be masked by the diffusion time 
of neutrinos ($\sim 1-10$ s)\cite{supernova} out of the dense medium and 
therefore probably indistinquishable from the first burst.

We have also studied the case of $T_p=0.1$~MeV (not shown in the
figures). The photon luminosity can be maintained at $10^{45}$ erg
s$^{-1}$ up to $5\times10^6$ s,
and it then decreases gradually to $10^{34}$ erg s$^{-1}$ in $10^{8-9}$ s,
while the neutrino luminosity first drops to $10^{34}$ erg
s$^{-1}$ and then rises to $10^{41}$ erg s$^{-1}$ during the mixed phase.
If such a long duration $\gamma$-ray source exists, it will be difficult
to be explained by other models. For our model with $T_p\lesssim0.01$~MeV
(not shown), the cooling history is similar, but with an even longer
duration and lower luminosities in the mixed phase.

The cooling curves (of all our models) are basically cooling of a bare
quark star plus that of a neutron star. When the quark star cools
to the phase transition temperature, the cooling curve switches from that of
the quark star to the neutron star, plus a phase transition epoch. We can
see that for a typical bare quark star without phase transition, the
photon luminosity, as well as the surface temperature, are higher than that
of the neutron star in the early epoch. The quark star then cools much
faster than the neutron star, and it will be too cold to be observed. For
the cooling of a bare
quark star with phase transition, there is a significant feature
different from the case without phase transition. When the quark star cools
to the phase transition temperature, it remains at the phase transition
temperature, and it releases the latent heat for a long time depending on
the latent heat and the total luminosity. The total luminosity actually
depends on the temperature. This means that the higher the phase transition
temperature, the larger the luminosity and hence the shorter the duration
of the phase transition. The latent heat depends on the choice of the EOS's
of both the quark and neutron matter, as well as the mass of the
compact star. Here we have chosen $\Lambda = 2.473 \mu$ for quark matter,
the HFV EOS for neutron matter and a compact star mass of $\sim
1.4M_{\odot}$. The latent heat is hence $2.08\times 10^{53}$ erg.

\subsection{Model c}

For Model c, the thermal properties are the same as Models a and b. In
this model, the pion emissivity is not considered, and the dominating
cooling process in the
early epoch is the $e^-e^+$ production. Hence the photon luminosities for
Model c in the early epoch is significantly lower, and the durations of
phase transition for high $T_p$ models are longer. For $T_p=1$ MeV, the
photon luminosity drops from $10^{53}$ to $10^{48}$ erg s$^{-1}$ and
can maintain such a high luminosity up to $10^4$ s. Such a model seems to 
be in good agreement with long duration
GRB's both in the energy and time scales.
For $T_p=10$ MeV, the cooling curves are similar to those of Model a and
b, except for the lower photon luminosity with a slight rise near the
end of the phase transition in Model c. The similarity in cooling curves is
due to
the same dominating cooling mechanism of neutrino emission at the high
temperature epoch in the quark and mixed phase, as well as the same
cooling behaviour after the phase transition.
We have again studied the case for $T_p=0.1$~MeV (not shown in the figure),
the photon luminosity can be maintained up to $10^{38-39}$ erg s$^{-1}$ 
for $10^{13-14}$ s. This model is similar to Page and Usov's work 
\cite{Usov2002}, but our work can maintain a high luminosity for much 
longer time. The late and
phase transition epoch of the cooling curve for $T_p\lesssim0.01$~MeV (not
shown in the figure) are
the same as those of Models a and b at the same $T_p$. It is because for
temperature $\lesssim 0.01$~MeV, the dominating cooling mechanisms in all
models (including Models d - e as well) in the quark and mixed phases
are non-equilibrium blackbody radiation.

\subsection{Model d, e}

As compared to Models a and b, the cooling curves for Models
d and e are also nearly identical. It is because the main
effect of a large superfluid gap is to suppress pion and neutrino
emissivities, which are important for the high temperature epoch. Hence
the $e^-e^+$
production dominates the photon luminosity in the early epoch. Thus both
models are dominantly cooled by the same mechanisms. We
can see that the effect of superfluidity is to make the neutrino bursts
more distinct. For $T_p=10$ MeV, the neutrino flux rises by over ten
orders of magnitudes within a small fraction of second. The effect may
again be masked by the diffusion of neutrinos. However, if $T_p$ is as
low as
1 MeV, the two bursts of similar flux can be separated by as long as $10^5$
s, which should be observable by modern neutrino observatories.

\subsection{Model f}

In Model f, $e^-e^+$ production is not considered. Although pion
emissivity is included, it is again greatly suppressed. Hence the
cooling history is as if there is no fast cooling mechanism. The quark
star cools slowly, maintaining a relatively low photon luminosity for
$\sim 10^8$ s for $T_p=1$ MeV. One special feature of this model is that
the photon luminosity rises up for about six orders of magnitudes at the
end of
the phase transition for $T_p=1$ MeV. This is due to the increase in
luminosity of thermal equilibrium radiation during the phase transition
[Section \ref{section:PT}]. Indeed for Model a - e, if we separate the
spectrum of thermal equilibrium radiation from other photon luminosities,
it is also increasing. In our
models, we expect more photons to be released during the phase
transition. The duration is $\sim 10$ s in this model and are shorter
for other models. If such a burst is observed, it will be a unique feature
which is difficult to be explained by other models.
We remark here that detailed radiative transfer calculation may kill or
lower the peak of this burst due to the surface
effect of the star.  Another special feature of this model is that for
all Models a - e (except Model c), the second neutrino burst starts within 
~1 s from the first burst, which may
be too short to be detected. This agrees with the observational data of
SN1987a \cite{supernova} (we emphasis here that base on the one 
event and the very few neutrinos detected, we really cannot tell 
how many neutrino bursts are there). 
In Model~f, the second neutrino burst starts after $10^{4-5}$ seconds. 
Such a delayed neutrino burst may be detectable, and it may have 
significant effects on the propagation of shock waves in supernova 
remnants \cite{supernova}.

\subsubsection{Remarks}
In these models, we can see that the size of superfluid gap determines
how a compact star cools in the early epoch. Pion emission is a very
efficient cooling mechanism if the superfluid gap is small. If the
superfluid gap is large, $e^-e^+$ production will dominate in the early
epoch. Also we find that a second neutrino burst is a signature of the phase
transition, if the neutrino emissivity in the neutron phase is higher than
that of the quark phase. In our models, the URCA process in a neutron star
makes the second neutrino burst possible.

\section{Discussion and conclusion}
We studied the global structure and stability of strange stars with 
the perturbative QCD EOS up to order $\alpha_s^2$. We find that for 
$\Lambda 
\geq 2.88 \mu$, strange quark matter is absolutely stable, while for 
$\Lambda \geq 2.35 (\gtrsim 2.7) \mu$, a strange star is globally 
stable (compared to a neutron star). 
The effect of 
strange quark mass to the strange star is also studied. For a strange 
quark mass $\leq 150(225)$ MeV, the correction to the maximum mass of 
strange stars is less than 3(10)\%. We suggest that a strange star may 
undergo a phase transition to a neutron star when it cools down to some 
temperature $T_p$.

It has been argued that the rapid cooling of strange stars by pion emission 
can be a power source of GRB \cite{Ng}. However, if 
the phase transition is not considered, the duration of the burst is too 
short, $<10^{-2}$ s, which cannot explain long duration GRB's, 
due to the rapid cooling without the maintenance of high temperature. 
However, in our model, the latent heat of the phase 
transition can supply the energy of the order $10^{53}$ erg. 
In our models, the photon luminosities can be maintained in 
$10^{48}-10^{54}$ erg s$^{-1}$ with duration from $\sim 10^{-3}$ up to 
$10^4$~s (or even longer), which are in the range for both long and 
short $\gamma$-ray bursts.
Thus the latent heat solves the problem of the energy supply for 
the GRB's. 

Another signature for quark stars that has been discussed is the 
relatively low luminosity of $\gamma$-rays when the quark star cools 
down \cite{Usov2002}. For models with low $T_p$, e.g. 
Model c with $T_p \sim 0.1$~MeV, the luminosity due to the $e^-e^+$ 
pairs in the $\gamma$-ray range can be as high as $10^{38-39}$ erg 
s$^{-1}$with a very long duration of $\sim 10^{13-14}$ s, which is much 
longer than the case considered by Page and Usov \cite{Usov2002}. The 
satellite $Integral$ launched recently is just sensitive to $\gamma$-rays 
in the energy range of 15 keV - 10 MeV. If long duration $\gamma$-ray 
source is really detected, it would be 
a signature of the phase transition. On the other hand, if no such source is 
detected, this does not mean failure of our models. Perhaps 
the phase transition temperature is very high, or the latent heat is small.
Indeed $Integral$ has already discovered a number of soft $\gamma$-ray 
point sources in the center of our galaxy \cite{lebrun}. Oaknin and 
Zhitnitsky have also discussed the possibility of the $\gamma$-ray 
sources being supermassive very dense droplets (strangelets) of dark 
matter in a recent paper \cite{oaknin}.

If a second neutrino burst after a supernova explosion is detected, 
it will be a 
strong evidence for the existence of quark star, and in addition, it will 
support the switch on of URCA process, or other fast neutrino emitting 
processes in neutron stars. However, the absence of the second burst does 
not kill quark stars, because there could simply be no fast neutrino 
emitting process in the neutron star phase, or the second burst arises so 
quickly that we cannot distinguish it from the first one.

It is generally believed that GRB's are related to 
supernovae \cite{sn-grb1,sn-grb2,sn-grb3}. Our model is compatible to the 
supernova 
connection --  a supernova explosion leaves behind a compact star, 
which triggers the GRB weeks to years later. We suggest that the phase 
transition from a hot strange star to a neutron star may be the central
engine of GRB's. During the phase 
transition, the size of the compact star increases, and an internal 
shock may develop, which initiates the GRB.

In our models, the phase transition is treated in an oversimplified manner. 
Detailed study should be made which involves hydrodynamics, the 
EOS, the emission properties of matter in the mixed phase and so on. 
Also the duration of the quark phase may be so short that for the cooling 
process, the hydrostatic treatment of quark stars may not be appropriate. 
Hydrodynamic simulation may be needed for the whole process starting from the
supernova explosion. This involves detailed numerical treatment which is 
beyond the scope of this paper. Here we discuss semi-quantitatively the 
signatures left during the phase transitions by the presence of the quark 
phase, if it is reached in supernovae.

This work is partially supported by a Hong Kong RGC Earmarked Grant 
CUHK4189/97P and a Chinese University Direct Grant 2060105.  We thank 
Prof.~K.~S.~Cheng for useful discussion.


\begin{table*}[h]
\begin{center}
\begin{tabular}{|c|c|c|c|c|c|}
\hline
EOS(NS) & $M_B/M_{\odot}$ &
${\mathcal{E}}_{\textrm{\scriptsize{binding}}}($NS$)$  & $\Lambda/\mu$ &
$M_G(SS)/M_{\odot}$ &
${\mathcal{E}}_{\textrm{\scriptsize{binding}}}($SS$)$
\\     &      & [MeV] &       &      & [MeV]
\\\hline
    HV & 1.51 & 68.4 & 2.473  & 1.44 & 43.5
\\  HV & 1.51 & 68.4 & 2.600  & 1.41 & 62.1
\\  HV & 1.51 & 68.4 & 2.880  & 1.33 & 111.9
\\  HV & 1.51 & 68.4 & 3.000  & 1.29 & 136.8
\\\hline
    HFV & 1.60 & 117.4 & 2.473  & 1.516 & 49.3
\\  HFV & 1.60 & 117.4 & 2.600  & 1.478 & 71.6
\\  HFV & 1.60 & 117.4 & 2.880  & 1.400 & 117.4
\\  HFV & 1.60 & 117.4 & 3.000  & 1.363 & 139.1
\\\hline

    $\Lambda^{\textrm{\scriptsize{RBHF}}}_{\textrm{\scriptsize{BroB}}}+$HFV$$
& 1.62 & 127.5  & 2.473 &   /  &   /
\\  $\Lambda^{\textrm{\scriptsize{RBHF}}}_{\textrm{\scriptsize{BroB}}}+$HFV$$
& 1.62 & 127.5  & 2.600 & 1.50 & 69.6
\\  $\Lambda^{\textrm{\scriptsize{RBHF}}}_{\textrm{\scriptsize{BroB}}}+$HFV$$
& 1.62 & 127.5  & 2.880 & 1.41 & 121.7
\\  $\Lambda^{\textrm{\scriptsize{RBHF}}}_{\textrm{\scriptsize{BroB}}}+$HFV$$
& 1.62 & 127.5  & 3.000 & 1.38 & 139.1
\\\hline
    $\textrm{G}^{\textrm{\scriptsize{K240}}}_{\textrm{\scriptsize{M78}}}$
& 1.56 & 96.3 & 2.473 & 1.48 & 48.2
\\  $\textrm{G}^{\textrm{\scriptsize{K240}}}_{\textrm{\scriptsize{M78}}}$
& 1.56 & 96.3 & 2.600 & 1.45 & 66.2
\\  $\textrm{G}^{\textrm{\scriptsize{K240}}}_{\textrm{\scriptsize{M78}}}$
& 1.56 & 96.3 & 2.880 & 1.37 & 114.4
\\  $\textrm{G}^{\textrm{\scriptsize{K240}}}_{\textrm{\scriptsize{M78}}}$
& 1.56 & 96.3 & 3.000 & 1.33 & 138.4
\\\hline

$\textrm{G}^{\textrm{\scriptsize{K240}}}_{\textrm{\scriptsize{M78}}}($NP$)$ &
 1.56 & 96.3 & 2.473 & 1.48 & 48.2
\\
$\textrm{G}^{\textrm{\scriptsize{K240}}}_{\textrm{\scriptsize{M78}}}($NP$)$ &
 1.56 & 96.3 & 2.600 & 1.45 & 66.2
\\
$\textrm{G}^{\textrm{\scriptsize{K240}}}_{\textrm{\scriptsize{M78}}}($NP$)$ &
 1.56 & 96.3 & 2.880 & 1.37 & 114.4
\\
$\textrm{G}^{\textrm{\scriptsize{K240}}}_{\textrm{\scriptsize{M78}}}($NP$)$ &
 1.56 & 96.3 & 3.000 & 1.33 & 138.4

\\\hline
\end{tabular}
\end{center}
\caption{\small
Binding energy, ${\mathcal{E}}_{\textrm{\scriptsize{binding}}}$, for
various compact stars.
A neutron star
gravitational mass $M_G({\rm NS})=1.4M_{\odot}$ is assumed, and
the baryonic masses of strange stars are chosen to be equal those of the
neutron stars. The many-body approximation for HV, HFV,
$\Lambda^{\textrm{\scriptsize{RBHF}}}_{\textrm{\scriptsize{BroB}}}+$
HFV
and
$\textrm{G}^{\textrm{\scriptsize{K240}}}_{\textrm{\scriptsize{M78}}}$
EOS's are relativistic Hartree, relativistic Hartree-Fock, relativistic
Brueckner-Hartree-Fock + relativistic Hartree-Fock and relativistic
Hartree respectively~[32].
}
\label{table:stability:binding}
\end{table*}

\newpage

\begin{table*}[H]
\caption{Effect of strange quark mass on the  global structure of
a strange star. The superscripts of `max' correspond to quantities of 
maximum mass stars.} 
\label{table:ms:ms_corr}
\begin{center}
\begin{tabular}{|c|c|c|c|c|c|}
\hline
\multicolumn{6}{|c|}{$\Lambda=2.473\mu$}\\
\hline
$m_s$(MeV) & $\epsilon_c^{\rm max}(\epsilon_0)$ & $M^{\rm max}(M_\odot)$
& $\%$ increase & $R^{\rm max}$ (km) & $\%$ increase          \\
\hline
  0   & 12.0 & 1.516 & $/$        & 8.54 & $/$
\\75  & 12.3 & 1.550 & $+2.22\%$  & 8.64 & $+1.17\%$
\\150 & 12.1 & 1.533 & $+1.11\%$  & 8.59 & $+0.59\%$
\\225 & 14.0 & 1.441 & $-4.95\%$  & 8.15 & $-4.57\%$
\\300 & 16.1 & 1.323 & $-12.75\%$ & 7.69 & $-9.95\%$
\\\hline
\end{tabular}
\end{center}
\begin{center}
\begin{tabular}{|c|c|c|c|c|c|}
\hline
\multicolumn{6}{|c|}{$\Lambda=2.88\mu$}\\
\hline
$m_s$(MeV) & $\epsilon_c^{\rm max}(\epsilon_0)$ & $M^{\rm max}(M_\odot)$
& $\%$ increase & $R^{\rm max}$ (km) & $\%$ increase          \\
\hline
  0   &  8.0 & 1.983 & $/$        & 11.07 & $/$
\\75  &  7.2 & 2.022 & $+1.93\%$  & 11.33 & $+2.35\%$
\\150 &  7.3 & 1.956 & $-1.36\%$  & 11.08 & $+0.09\%$
\\225 &  8.8 & 1.796 & $-9.46\%$  & 10.35 & $-6.50\%$
\\300 &  9.5 & 1.641 & $-17.26\%$ &  9.92 & $-10.39\%$
\\\hline
\end{tabular}
\end{center}
\end{table*}

\newpage

\begin{table*}[H]
\caption{Total conversion energy $E_{\textrm{\scriptsize{conv}}}$
released of a compact star for different $\Lambda$. The neutron star 
gravitational mass $M_G=1.4M_{\odot}$ with various EOS's.}
\label{table:cooling:conv}
\begin{center}
\begin{tabular}{|c|c|c|c|c|}
\hline
EOS(NS) & $M_B/M_{\odot}$ & $\Lambda/\mu$ & $M_G(SS)/M_{\odot}$ &
$E_{\textrm{\scriptsize{conv}}}/10^{53}$ erg
\\\hline
    HV & 1.51 & 2.473  & 1.44 & $+0.72$
\\  HV & 1.51 & 2.600  & 1.41 & $+0.18$
\\  HV & 1.51 & 2.880  & 1.33 & $-1.25$
\\  HV & 1.51 & 3.000  & 1.29 & $-1.97$
\\\hline
    HFV & 1.60 & 2.473  & 1.516 & $+2.08$
\\  HFV & 1.60 & 2.600  & 1.478 & $+1.40$
\\  HFV & 1.60 & 2.880  & 1.400 & 0
\\  HFV & 1.60 & 3.000  & 1.363 & $-0.66$
\\\hline
    
    $\Lambda^{\textrm{\scriptsize{RBHF}}}_{\textrm{\scriptsize{BroB}}}+$HFV$$ 
& 1.62 & 2.473  &   /  &   /     
\\  $\Lambda^{\textrm{\scriptsize{RBHF}}}_{\textrm{\scriptsize{BroB}}}+$HFV$$ 
& 1.62 & 2.600  & 1.50 & $+1.79$ 
\\  $\Lambda^{\textrm{\scriptsize{RBHF}}}_{\textrm{\scriptsize{BroB}}}+$HFV$$ 
& 1.62 & 2.880  & 1.41 & $+0.18$
\\  $\Lambda^{\textrm{\scriptsize{RBHF}}}_{\textrm{\scriptsize{BroB}}}+$HFV$$ 
& 1.62 & 3.000  & 1.38 & $-0.36$
\\\hline
    $\textrm{G}^{\textrm{\scriptsize{K240}}}_{\textrm{\scriptsize{M78}}}$ 
& 1.56 & 2.473  & 1.48 & $+1.43$ 
\\  $\textrm{G}^{\textrm{\scriptsize{K240}}}_{\textrm{\scriptsize{M78}}}$ 
& 1.56 & 2.600  & 1.45 & $+0.90$
\\  $\textrm{G}^{\textrm{\scriptsize{K240}}}_{\textrm{\scriptsize{M78}}}$ 
& 1.56 & 2.880  & 1.37 & $-0.54$ 
\\  $\textrm{G}^{\textrm{\scriptsize{K240}}}_{\textrm{\scriptsize{M78}}}$ 
& 1.56 & 3.000  & 1.33 & $-1.25$
\\\hline
    $\textrm{G}^{\textrm{\scriptsize{K240}}}_{\textrm{\scriptsize{M78}}}($NP$)$ & 1.56 & 2.473  & 1.48 & $+1.43$
\\  $\textrm{G}^{\textrm{\scriptsize{K240}}}_{\textrm{\scriptsize{M78}}}($NP$)$ & 1.56 & 2.600  & 1.45 & $+0.90$
\\  $\textrm{G}^{\textrm{\scriptsize{K240}}}_{\textrm{\scriptsize{M78}}}($NP$)$ & 1.56 & 2.880  & 1.37 & $-0.54$
\\  $\textrm{G}^{\textrm{\scriptsize{K240}}}_{\textrm{\scriptsize{M78}}}($NP$)$ & 1.56 & 3.000  & 1.33 & $-1.25$

\\\hline
\end{tabular}
\end{center}
\end{table*}

\newpage

\begin{table*}[H]
\caption{Models of quark phase presented in this paper.}
\label{model}
\begin{center}
\begin{tabular}{|c|p{1.5cm}p{1.5cm}p{1.5cm}p{1.5cm}p{1.5cm}|c|}
\hline
   &\multicolumn{5}{|c|}{Cooling Mechanisms} & Superfluid Gap \\
Model & quark URCA & neq & eq & $e^+e^-$ & pion & $\Delta$(MeV)
\\\hline
   a  & $\surd$ & $\surd$ & $\surd$ & $\times$ & $\surd$  & 1
\\ b  & $\surd$ & $\surd$ & $\surd$ & $\surd$  & $\surd$  & 1
\\ c  & $\surd$ & $\surd$ & $\surd$ & $\surd$  & $\times$ & 1
\\\hline
   d  & $\surd$ & $\surd$ & $\surd$ & $\surd$  & $\times$ & 100
\\ e  & $\surd$ & $\surd$ & $\surd$ & $\surd$  & $\surd$  & 100
\\ f  & $\surd$ & $\surd$ & $\surd$ & $\times$ & $\surd$  & 100
\\\hline
\end{tabular}
\end{center}
\end{table*}

\newpage

\newpage

\begin{figure}[h]
\caption{Cooling curves corrsponding to different models. Fig.~8.1 
corresponds to Models a and b. Fig.~8.2 corresponds to Model c. Fig.~8.3 
corrsponds to Model d and e. Fig.~8.4 corresponds to Model f. 
In each figure, the left
(right) panels are for models with small (large) phase transition
temperature $T_p$ = 1 (10) MeV. The dotted (dashed) lines are the cooling
curves of a pure strange (neutron) star without phase transition.
The solid line represents the scenario with phase transition.
$T_s^\infty, \ L_\gamma^\infty$ and $L_\nu^\infty$
denote surface temperature, photon luminosity and neutrino luminosity
respectively.  An initial temperature of 40 MeV is assumed. 
} 
\end{figure}








\newpage
\epsfig{file=cooling_result_model_cheng_a.eps,angle=0,width=8cm}
\centerline{FIG. 8.1}

\newpage
\epsfig{file=cooling_result_model_cheng_c.eps,angle=0,width=8cm}
\centerline{FIG. 8.2}

\newpage
\epsfig{file=cooling_result_model_cheng_d.eps,angle=0,width=8cm}
\centerline{FIG. 8.3}

\newpage
\epsfig{file=cooling_result_model_cheng_f.eps,angle=0,width=8cm}
\centerline{FIG. 8.4}


\begin{references}
\bibitem{Witten} E.~Witten, {\em Phys.~Rev.~D} {\bf 30}, 272 (1984).

\bibitem{Slane} P.~Slane, D.~J.~Helfand, and S.~S.~Murray,
{\em  Astrophys.~J.}, {\bf 571} L45-L49 (2002).

\bibitem{Tsuruta2002} S.~Tsuruta {\it et al.}, 
{\em  Astrophys.~J.}, {\bf 571} L143 (2002). 

\bibitem{Drake2002} J.~J.~Drake, H.~L.~Marshall, S.~Dreizler, 
P.~E.~Freeman, A.~Fruscione, M.~Juda, V.~Kashyap, F.~Nicastro, 
D.~O.~Pease, B.~J.~Wargelin, and K.~Werner
{\em Astrophys.~J.}~{\bf 572}, 996-1001 (2002).

\bibitem{Drake2003} S.~Zane, R.~Turolla, and J.~J.~Drake,
 {\em astro-ph/0302197}.

\bibitem{Alcock} C.~Alcock, E.~Farhi and A.~Olinto, {\em 
Astrophys.~J.}~{\bf 310}, 
261 (1986); P.~Haensel, J.~L.~Zdunik and R.~Schaeffer, 
Astron.~Astrophys.~{\bf 160}, 121 (1986). 

\bibitem{Glendenning} N.~K.~Glendenning, {\it Compact Stars} 
 (Springer, 1996).

\bibitem{Ng} C.~Y.~NG, K.~S.~Cheng, and M.~-C.~Chu, 
Astropart.~Phys.~{\bf 19}, 171-192 (2003).

\bibitem{Usov1998} V.~V.~Usov, {\em Phys.~Rev.~Lett.} {\bf 80}, 230 (1998).

\bibitem{Usov2001} V.~V.~Usov, {\em  Astrophys.~J.}, {\bf 550} 
L179-L182 (2001). 

\bibitem{Usov2002} D.~Page and V.~V.~Usov, 
{\em Phys.~Rev.~Lett.} {\bf 89}, 131101 (2002).

\bibitem{shuryak} See for example, E.~V.~Shuryak, {\it The QCD vacuum, 
hadrons and the superdense matter} (World Scientific, Singapore, 1988) and
references therein.

\bibitem{kawa1} K.~-W.~Wong and M.~-C.~Chu, MNRAS {\bf 350}, L42 (2004).

\bibitem{supernova} Albert G. Petschek, {\it Supernovae}, (Springer, 1990).

\bibitem{fodor} Z.~Fodor and S.~D.~Katz, JHEP {\bf 0203}, 014 (2002).

\bibitem{allton} C.~R.~Allton {\it et al.}, Phys.~Rev.~D {\bf 66},  
074507 (2002).

\bibitem{Benvenuto} O.~G.~Benvenuto and G.~Lugones, MNRAS {\bf 304}, L25
(1999).

\bibitem{lebrun} F.~Lebrun{\it et al.},
 {\em Nature}~{\bf 428}, 293 (2004).

\bibitem{oaknin} D.~H.~Oaknin, and A.~R.~Zhitnitsky, hep-ph/0406146.

\bibitem{Aguilera} D.~N.~Aguilera, D.~Blaschke, and H.~Grigorian,
Astron.~Astrophys.~{\bf 416}, 991 (2004). 

\bibitem{Fraga} E.~S.~Fraga, R.~D.~Pisarski, and J.~Schaffner-Bielich, 
 {\em Phys.~Rev.~D} {\bf 63}, 121702 (2001).

\bibitem{Freedman} B.~Freedman, and L.~McLerran,  
 {\em Phys.~Rev.~D} {\bf 17}, 1109-1122 (1978).

\bibitem{Baluni} V.~Baluni,  
 {\em Phys.~Rev.~D} {\bf 17}, 2092 (1978).



\bibitem{Farhi} E.~Farhi and R.~L.~Jaffe, {\em Phys.~Rev.~D} {\bf 30}, 
2379-2390 (1984).

\bibitem{kawa-thesis} K.~W.~Wong, Chinese University MPhil 
Thesis (2003), unpublished.

\bibitem{Rajagopal} K.~Rajagopal, and F.~Wilczek,  
 {\em Phys.~Rev.~Lett.}~{\bf 86}, 3492 (2001).

\bibitem{Peng} G.~X.~Peng, H.~C.~Chiang, P.~Z.~Ning, and B.~S.~Zhou,  
 {\em Phys.~Rev.~C} {\bf 59}, 3452-3454 (1999).

\bibitem{Madsen} J.~Madsen,  {\em Phys.~Rev.~Lett.}~{\bf 85}, 4687-4690 
(2000).

\bibitem{cheng1} K.~S.~Cheng, and Z.~G.~Dai,
 {\em  Phys.~Rev.~Lett.}~{\bf 77} 1210-1213 (1996). 

\bibitem{cheng2} K.~S.~Cheng, and Z.~G.~Dai,
 {\em  Phys.~Rev.~Lett.}~{\bf 80} 18 (1998). 


\bibitem{Bombaci} I.~Bombaci, and B.~Datta,
 {\em  Astrophys.~J.}, {\bf 530} L69-L72 (2000). 

\bibitem{Weber} F.~Weber, {\it Pulsars as Astrophysical Laboratories 
for Nuclear and Particle Physics}, (Institute of Physics, Bristol, 1999).

\bibitem{phillips} A.~C.~Phillips, {\it The Physics of Stars}, (John 
Wiley and Sons, 1994).

\bibitem{Srivastava} D.~K.~Srivastava and B.~Sinha,  
 {\em Phys.~Rev.~Lett.~} {\bf 73}, 2421-2424 (1994).









\bibitem{Gudmundsson} E.~H.~Gudmundsson, C.~J.~Pethick, and R.~I.~Epstein,
 {\em  Astrophys.~J.}~{\bf 272} 286-300 (1983). 

\bibitem{Iwamoto} N.~Iwamoto,  
 {\em Phys.~Rev.~Lett.}~{\bf 44}, 1637 (1980).

\bibitem{rebhan} A.~Ipp, A.~Gerhold and A.~Rebhan calculated the 
non-Fermi-liquid behavior of the specific heat, which is larger 
than that of the ideal case [{\em Phys.~Rev.~D} {\bf 69}, 
011901 (2004)]. If this is the case, the cooling of quark star will be 
slower.

\bibitem{Horvath} J.~E.~Horvath, O.~G.~Benvenuto, and H.~Vucetich,  
 {\em Phys.~Rev.~D} {\bf 44}, 3797 (1991).

\bibitem{Maxwell} O.~V.~Maxwell,
 {\em  Astrophys.~J.}~{\bf 231} 201-210 (1979). 

\bibitem{Csernai} L.~P.~Csernai, {\it Introduction to relativistic heavy 
ion collisions}, (Wiley, 1994).

\bibitem{Muller} B.~Muller, and J.~M.~Eisenberg,  
 {\em Nucl.~Phys.~{\bf A435}, 791 (1985)}.

\bibitem{cheng2003} K.~S.~Cheng, and T.~Harko,
 {\em  Astrophys.~J.}~{\bf 596} 451-463 (2003). 

\bibitem{Lattimer} J.~M.~Lattimer, C.~J.~Pethick, M.~Prakash, and 
P.~Haensel
 {\em Phys.~Rev.~Lett.}~{\bf 66}, 2701 (1991).

\bibitem{Festa??} G.~G.~Festa, and M.~A.~Ruderman,  
 {\em Phys.~Rev.}~{\bf 180}, 1227 (1969).

\bibitem{Lindblom} L.~Lindblom, B.~J.~Owen, and G.~Ushomirsky,  
 {\em Phys.~Rev.~D} {\bf 62}, 084030 (2000).

\bibitem{Gentile} N.~A.~Gentile, M.~B.~Aufderheide, G.~J.~Mathews, 
F.~D.~Swesty, and G.~M.~Fuller,
 {\em  Astrophys.~J.}~{\bf 414} 701-711 (1993). 

\bibitem{aksenov1} A.~G.~Aksenov, M.~Milgrom, and V.~V.~Usov , MNRAS {\bf 
343}, L69 (2003).

\bibitem{aksenov2} A.~G.~Aksenov, M.~Milgrom, and V.~V.~Usov, 
 {\em  Astrophys.~J.}~{\bf 609} 363-377 (2004). 

\bibitem{Tsuruta1998} S.~Tsuruta,  
 {\em Phys.~Rep} {\bf 292}, 1 (1998).

\bibitem{Hirata} K.~Hirata {\it et al.},
 {\em Phys.~Rev.~Lett.}~{\bf 58}, 1490 (1987).


\bibitem{sn-grb1} J.~S.~Bloom {\it et al.},
 {\em Nature}~{\bf 401}, 453 (1999).

\bibitem{sn-grb2} J.~N.~Reeves {\it et al.},
 {\em Nature}~{\bf 416}, 512 (2002).

\bibitem{sn-grb3} E.~Waxman {\it et al.},
 {\em Nature}~{\bf 423}, 388 (2003).



\end{references}
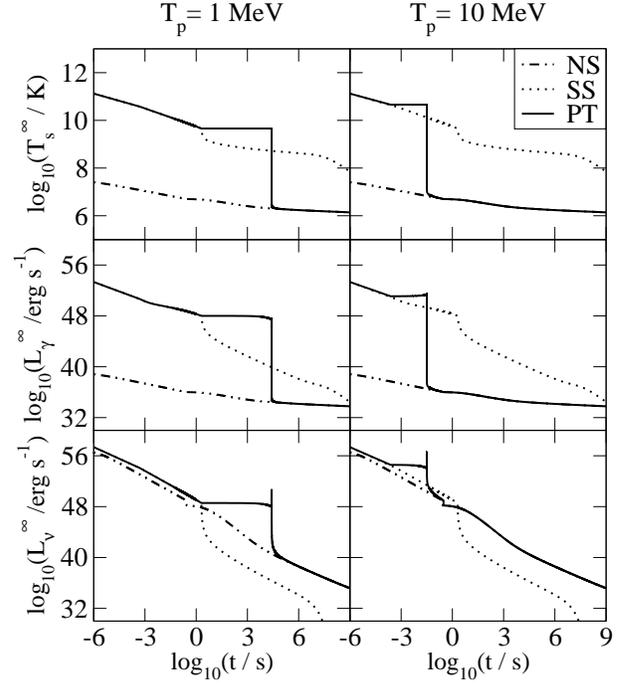
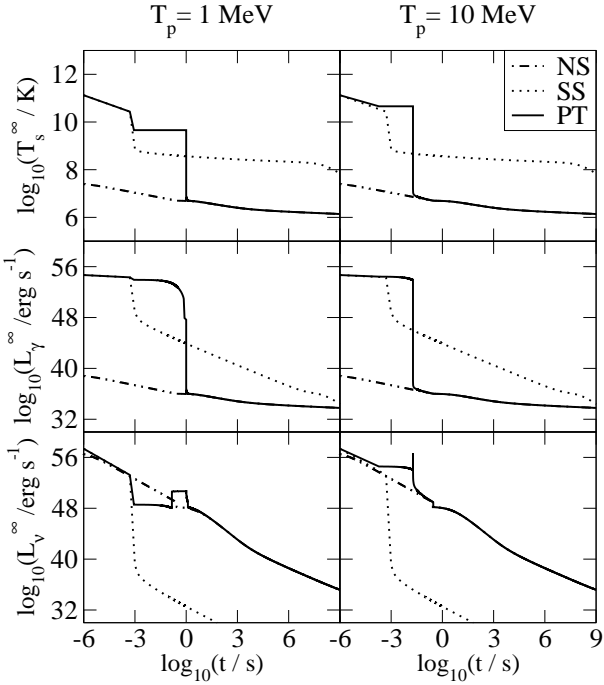
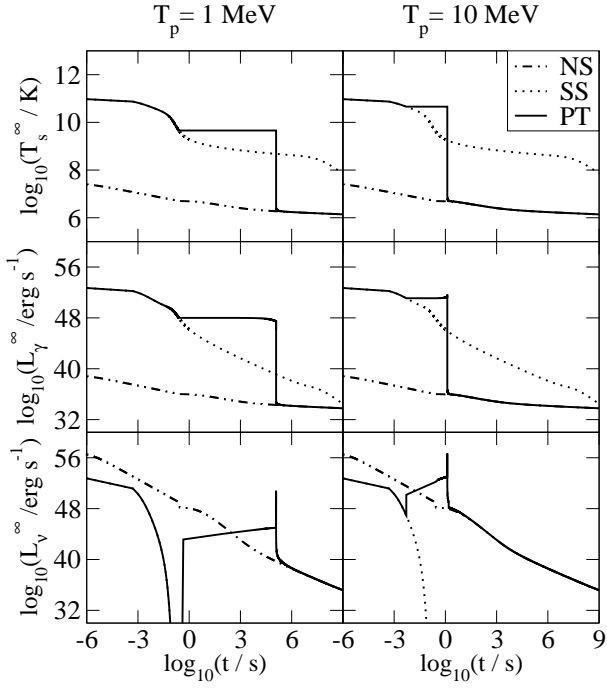
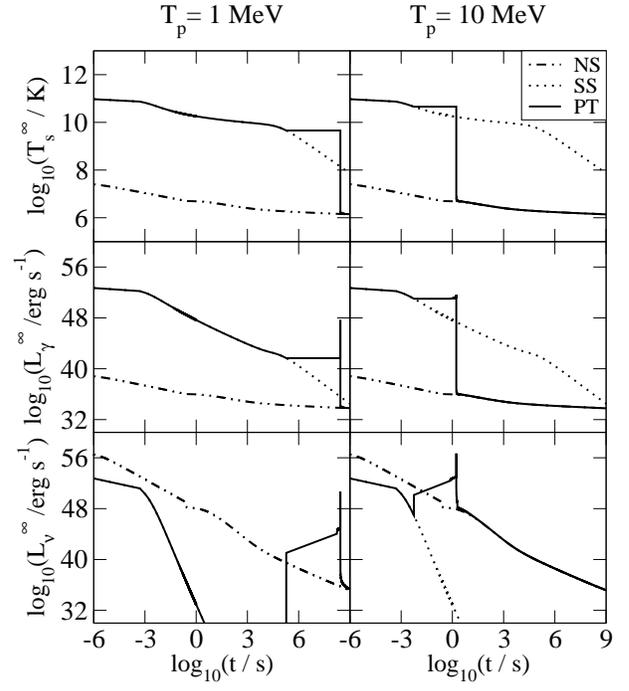
\end{document}